\documentclass[twocolumn]{aastex61}
\usepackage{amsmath,amstext}
\usepackage[figure,figure*]{hypcap}
\usepackage{multirow}
\usepackage{graphicx,url,pifont}
\usepackage{amssymb,color}

\def\gsim{\mathrel{\hbox{\rlap{\hbox{\lower4pt\hbox{$\sim$}}}\hbox{$>$}}}}

\newcommand{\Apostle}{\textsc{Apostle}}
\newcommand{\Auriga}{\textsc{Auriga}}

\newcommand{\Gadget}{\textsc{Gadget-2}}
\newcommand{\PGadget}{\textsc{Gadget-3}}

\newcommand{\Subfind}{\textsc{Subfind}}

\newcommand{\Galform}{\textsc{Galform}}
\newcommand{\coco}{\textsc{coco}}
\newcommand{\Color}{\textsc{color}}
\newcommand{\zcut}{z_{{\rm cut}}}
\newcommand{\vcut}{{\rm V}_{{\rm cut}}}

\newcommand{\bq}{\begin{eqnarray}}
\newcommand{\eq}{\end{eqnarray}}

\submitjournal{ApJ}
\shorttitle{Luminosity functions and reionisation}
\shortauthors{S. Bose {\it et al.}}

\begin{document}

\title{The imprint of cosmic reionisation on the luminosity function of galaxies}
\correspondingauthor{Sownak Bose}
\email{sownak.bose@cfa.harvard.edu}

\author[0000-0002-0974-5266]{Sownak Bose}
\affil{Harvard-Smithsonian Center for Astrophysics, 60 Garden St., Cambridge, MA 02138, USA} 

\author[0000-0001-6146-2645]{Alis J. Deason}
\affil{Institute for Computational Cosmology, Durham University, South Road, Durham, DH1 3LE, UK}

\author[0000-0002-2338-716X]{Carlos S. Frenk}
\affil{Institute for Computational Cosmology, Durham University, South Road, Durham, DH1 3LE, UK}

\begin{abstract}
  The (re)ionisation of hydrogen in the early universe has a profound
  effect on the formation of the first galaxies: by raising the gas
  temperature and pressure, it prevents gas from cooling into small
  haloes thus affecting the abundance of present-day small
  galaxies. Using the \Galform{} semi-analytic model of galaxy
  formation, we show that two key aspects of the reionisation process
  -- {\it when} reionisation takes place and the {\it characteristic
    scale} below which it suppresses galaxy formation -- are imprinted
  in the luminosity function of dwarf galaxies. We focus on the
  luminosity function of satellites of galaxies like the Milky Way and
  the LMC, which is easier to measure than the luminosity function of
  the dwarf population as a whole. Our results show that the details
  of these two characteristic properties of reionisation determine the
  {\it shape} of the luminosity distribution of satellites in a unique
  way, and is largely independent of the other details of the galaxy
  formation model. Our models generically predict a bimodality in the
  distribution of satellites as a function of luminosity: a population
  of faint satellites and population of bright satellites separated by
  a `valley' forged by reionisation. We show that this bimodal
  distribution is present at high statistical significance in the
  combined satellite luminosity function of the Milky Way and M31. We
  make predictions for the expected number of satellites around
  LMC-mass dwarfs where the bimodality may also be measurable in
  future observational programmes. Our preferred model predicts a
  total of $26 \pm 10$ (68 per cent confidence) satellites brighter
  than ${\rm M}_V=0$ in LMC-mass systems.
\end{abstract}

\keywords{
   galaxies: luminosity function -- galaxies: dwarf -- galaxies: formation --
   cosmology: reionisation
}

\section{Introduction}
\label{sect:Intro}

In the $\Lambda$CDM model of structure formation, small dark matter
haloes are already forming profusely during the {\it dark ages} -- the
period following the (re)combination of hydrogen at redshift, $z \sim
1100$, when the gas becomes neutral. Neutral gas is able to cool into
these dark matter haloes and form stars and galaxies, bringing the
dark ages to an end. UV radiation from the first sources of light
reionises the hydrogen heating it up to a temperature of $\sim
10^4$~K, raising its entropy, and preventing it from cooling into
haloes with effective temperature, $T_{{\rm vir}} \lesssim 10^4$~K
\citep{Doroshkevich1967, Couchman1986}. Thus, the reionisation process
temporarily halts the formation of galaxies in low mass haloes
\citep{Rees1986,Efstathiou1992,Loeb2001}. Galaxy formation resumes
some time later when sufficiently massive haloes, with virial
temperature well above $10^4$~K, begin to form.

The temporary suppression of galaxy formation as a result of
reionisation is reflected in the abundance of dwarf galaxies today. No
galaxies form below a present-day halo mass of a few times
$10^7M_\odot$, and only a fraction of the haloes with mass between
this value and $\sim 10^{10}M_\odot$ form a galaxy
\citep{Sawala_2013,Sawala_2016a,Fitts_2017}. In haloes that do form a
galaxy, the growth of stellar mass is further limited by supernovae
feedback \citep{Larson_1974,White1978,White1991}. One consequence of
these processes is that the number of these `chosen few' galaxies is
much smaller than the number of dark matter subhaloes predicted to be
orbiting around the Milky Way in cosmological $N$-body simulations
\citep[e.g.][]{Kauffmann1993,Bullock2000,Benson2002a,Benson2002b,Somerville2002,Font2011,Sawala2016},
thus readily explaining away the so-called {\it ``missing satellites
  problem''} often deemed to afflict the $\Lambda$CDM model
\citep{Klypin1999,Moore1999}.

The two critical features of reionisation that impact on the abundance
of dwarf galaxies are:
\begin{enumerate}
\item The {\it time} when reionisation happened.
\item {\it The characteristic scale} below which gas could no longer
  cool in dark matter haloes.
\end{enumerate}
These two features are linked because the epoch of reionisation
determines how long small mass haloes are able to continue forming
stars before reionisation inhibits further gas cooling in
them. Understanding these features is therefore crucial to an
understanding of galaxy formation.

An important constraint on the epoch of reionisation can be derived
from the polarisation of the cosmic microwave background (CMB)
radiation. The polarisation data directly constrain the electron
scattering optical depth to recombination, $\tau$, which can be
converted to an equivalent redshift of reionisation by assuming a
model for the redshift evolution of the ionisation fraction. Recent
estimates from the {\it Planck} satellite data imply that the Universe
was 50\% ionised by $z_{re}=8.8^{+1.7}_{-1.4}$
\citep{Planck2016}. Theoretical work suggests that reionisation
proceeded relatively quickly, with the ionisation fraction increasing
from 20\% to 90\% over $\sim 400$ Myr between $9\lesssim z \lesssim 6$
\citep{Robertson2015,Sharma2017}. An alternative probe of the
ionisation state of the IGM comes from the spectra of QSOs: the
absence of a Gunn-Peterson trough in the absorption spectra of quasars
at $z \lesssim 6$ \citep[e.g.][]{Fan2000}, and its presence in spectra
at $z \gtrsim 6$ \citep[e.g.][]{Becker2001,Bolton2011}, suggest that
the Universe completed the transition from neutral to ionised at
around that time. A third source of evidence is the decline in
Ly-$\alpha$ emission observed from galaxies at $z>6$, attributed to
absorption by intervening HI gas
\citep{Stark2006,Fontana2010,Caruana2012,Treu2013,Tilvi2014,Schenker2014,Caruana2014,Pentericci2014,Mason2017}.

There is still considerable uncertainty regarding the characteristic
scale below which galaxies are significantly affected by the
photoionising background. \cite{Rees1986} suggested that haloes with
circular velocities $\sim 30$ kms$^{-1}$ would be able to confine
photoheated gas in stable equilibrium (i.e., with photoheating
balanced by radiative cooling), an idea recently corroborated by
\cite{Benitez-Llambay_2017} in the \Apostle{} hydrodynamical
simulations \citep{Fattahi2016,Sawala2016}. \cite{Gnedin2000}
expressed the characteristic scale in terms of a filtering mass
(corresponding to a circular velocity of $\sim 50$ kms$^{-1}$) that
sets the scale over which baryonic perturbations are smoothed over in
linear perturbation theory \citep[see also][who reached a similar
  conclusion using 1D hydrodynamical
  simulations]{Thoul1996}. \cite{Okamoto2008} used high resolution
hydrodynamical simulations to estimate the loss of baryons from low
mass haloes resulting from photoionisation and revised the filtering
mass scale down to $\sim 25$ kms$^{-1}$ (corresponding to a halo mass
of $\sim 6 \times 10^9 M_\odot$). Recent radiation-hydrodynamic
simulations of reionisation give a halo mass of $\sim 2 \times
10^{9}\,M_\odot$ below which the effects of photoionisation become
important \citep{Ocvirk2016}.

In this paper we propose a new probe of the physics of reionisation:
the {\it shape} of the (differential) dwarf galaxy luminosity
function. We show explicitly that this function encodes both {\it
  when} reionisation happened and the {\it characteristic scale} below
which it had a significant impact. We focus specifically on the
luminosity function of satellites of both Milky Way and LMC-mass
galaxies because these are easier to measure observationally than the
luminosity function of the dwarf galaxy population as a
whole. However, all the features of the satellite luminosity functions
that we highlight here are also present in the general dwarf galaxy
luminosity function. Current observational surveys like the SDSS
\citep{Adelman2007,Alam2015}, the Dark Energy Survey
\citep[DES,][]{Bechtol2015,Drlica2015,Kim2015,Koposov2015} and
Pan-STARRS1 \citep{Laevens2015,Chambers2016} are rapidly improving the
census of faint satellites in the Milky Way.  The {\it total}
luminosity function of Milky Way satellites can be readily inferred
from a partial census
\citep{Koposov2008,Tollerud2008,Hargis2014,Newton2017}. Future surveys
like DESI and LSST will measure properties for large samples that may
enable estimates of the luminosity function of the dwarf galaxy
population as a whole and of satellites of LMC-mass galaxies which
will test the ideas developed in this paper.

This paper is organised as follows. In \S\ref{sect:numerical}, we
introduce the theoretical aspects of this work, including a systematic
investigation of how reionisation shapes the luminosity function of
satellites (\S\ref{sect:reionLF}). In \S\ref{sect:galactic}, we
combine the satellite populations of the Milky Way and M31 to test if
the imprint of reionisation can be detected in the observed luminosity
function of dwarf galaxies. In \S\ref{sect:lumMW}, we present the
cumulative and differential luminosity functions of the Milky Way
satellites predicted by our models and compare them to the data. We
also provide the predictions of our models for the satellite
luminosity function of LMC-mass haloes
(\S\ref{sect:predLMC}). Finally, our conclusions are summarised in
\S\ref{sect:Conclusions}.

\section{Theoretical considerations}
\label{sect:numerical}

In this section we provide an overview of the $N$-body simulations
used in this work. We describe the semi-analytic model of galaxy
formation, \Galform{}, used to populate dark matter haloes in the
simulation with galaxies. We also explore how reionisation shapes the
luminosity function of dwarf satellite galaxies.

\subsection{The {\it Copernicus Complexio} simulations}
\label{sect:coco}

The $N$-body simulations studied in this paper are part of the {\it
  Copernicus Complexio} (\coco{}) suite of simulations introduced by
\cite{Hellwing2016} and \cite{Bose2016}. \coco{} is a set of
cosmological zoom-in simulations that follow about 12 billion high
resolution dark matter particles, each of mass $m_p = 1.61 \times
10^5\,M_\odot$. The re-simulation region (roughly 24 Mpc in radius)
was extracted from the (100 Mpc)$^3$ parent volume, {\it Copernicus
  complexio Low Resolution} (\Color{}). Both \coco{} and \Color{}
assume cosmological parameters derived from the 7-yr {\it Wilkinson
  Microwave Anisotropy Probe} (WMAP-7) data \citep{Komatsu2011}:
$\Omega_m = 0.272$, $\Omega_\Lambda = 0.728$ and $h = 0.704$, where
$h$ is related to the present-day Hubble constant, $H_0$, by $h =
H_0/100{\rm kms}^{-1}{\rm Mpc}^{-1}$. The spectral index of the
primordial power spectrum is $n_s = 0.967$, and the linear power
spectrum is normalised at $z=0$ taking $\sigma_8 = 0.81$.

\coco{} was evolved from $z=127$ to $z=0$ using the \PGadget{} code,
an updated version of the publicly-available \Gadget{} code
\citep{Springel2001,Springel2005}. \coco{} consists of two sets of
simulations: one where the dark matter is CDM, and another where the
dark matter is a thermal relic warm dark matter (WDM) particle with a
rest mass of 3.3 keV; in this paper, we use only the CDM
simulation. For a more detailed description of the initial conditions
and re-simulation strategy in \coco{}, we refer the reader to
\cite{Hellwing2016} and \cite{Bose2016}.

Dark matter haloes were identified using the friends-of-friends
algorithm \citep{Davis1985}, while their self-bound substructures were
subsequently identified using the \Subfind{} algorithm
\citep{Springel2001b}. By requiring convergence of the mass function
in \coco{} with that obtained from its lower-resolution parent
simulation, \Color{}, we determine the resolution limit of our
simulations to be 300 dark matter particles, or $\sim 4.8 \times
10^7\,M_\odot$ in halo mass.

In this paper we are interested in the luminosity function of
satellites residing in Milky Way-mass and LMC-mass host haloes. In
what follows, a Milky Way-mass host is defined as a halo of mass at
$z=0$ in the range $7 \times 10^{11} \leq M_{200}/M_\odot \leq 2
\times 10^{12}$ \citep[e.g.][see \citealt{Wang2015} for a
  comprehensive list of
  references]{Smith2007,Deason2012,BoylanKolchin2013}; an LMC-mass
host is defined as a halo of mass in the range $1.5 \times 10^{11}
\leq M_{200}/M_\odot \leq 3.5 \times 10^{11}$ \citep[e.g.][Cautun et
  al. in prep.]{Besla2012,Besla2015,Penarrubia2016}. Here, $M_{200}$
is the mass contained within the virial radius, $r_{200}$, the radius
that encloses a mean density equal to 200 times the critical density
of the Universe at a given redshift. Using these criteria, we identify
85 Milky Way-mass hosts and 292 LMC-mass hosts at $z=0$ in
\coco{}. The merger trees from \coco{} are populated with galaxies
using a semi-analytic model of galaxy formation, which we now
describe.

\newpage

\subsection{The \Galform{} semi-analytic model of galaxy formation}
\label{sect:galform}

\subsubsection{The semi-analytic philosophy}
\label{sect:philosophy}

Semi-analytic models of galaxy formation are very instructive
theoretical tools for understanding the physics of galaxy
formation. While, by assuming spherical symmetry, semi-analytic models
are unable to follow the evolution of gas in galaxies in full
generality, as is done in hydrodynamical simulations, they are much
cheaper computationally. A great advantage of this is that, in
addition to generating a large, statistical sample of galaxies, it is
possible to explore rapidly the parameter space describing the
physical processes implemented in the model. This makes it
straightforward to examine how model predictions are affected by
turning on or off particular mechanisms; this is a feature we exploit
in \S\ref{sect:reionLF}. For further discussion of the methodology and
philosophy behind semi-analytic modelling, we refer the reader to the
review by \cite{Baugh2006}.

The Durham semi-analytic model of galaxy formation, \Galform{}, was
first presented in \cite{Cole1994} and~\cite{Cole2000}; it
incorporates the various physical processes thought to be important
for galaxy formation, such as the cooling of gas in haloes; star
formation in galactic disks and central starbursts; metal enrichment
of the interstellar medium (ISM); chemical evolution of stellar
populations; feedback from stellar winds, supernovae and active
galactic nuclei (AGN). As the demands on galaxy formation models have
increased due to the availability of better observational data, the
\Galform{} model has undergone several upgrades. For example,
\cite{Baugh2005} introduced a top-heavy IMF in starbursts in order to
reproduce the observed number counts of submillimetre galaxies. To
explain the exponential tail at the bright end of the galaxy
luminosity function, \cite{Bower2006} introduced AGN feedback as a
means to suppress star formation in bright galaxies. Motivated by the
improved observational understanding of the link between star
formation rates and the gas content of galaxies, \cite{Lagos2011}
introduced a star formation law that depends on the molecular gas
content of the ISM. \Galform{} employs the \cite{Maraston2005} stellar
population synthesis model to compute broad-band luminosities and
magnitudes from the stellar SEDs of galaxies.

In \Galform{} (and, to some extent, in some of the other semi-analytic
models currently in use,
e.g. \citealt{Menci2002,Monaco2007,Somerville2008,Guo2011,Benson2012,Henriques2015})
the free parameters of the model are set by requiring a good match to
a small selection of properties of the local galaxy population, in
particular for \Galform{}: (1) the optical and near-IR luminosity
functions at $z = 0$; (2) the HI mass function at $z=0$; (3) galaxy
morphological fractions at $z= 0$; (4) the normalisation of the black
hole -- bulge mass relation at $z=0$. \cite[A comprehensive list may
  be found in \S4.2 of][L16 hereafter]{Lacey2016}. In this sense,
while semi-analytic models do contain free parameters, the degree to
which they can be `tuned' is limited by demanding that a small set of
observational data be always reproduced.

\begin{figure*}[t!]
  \center{\includegraphics[width=\textwidth]{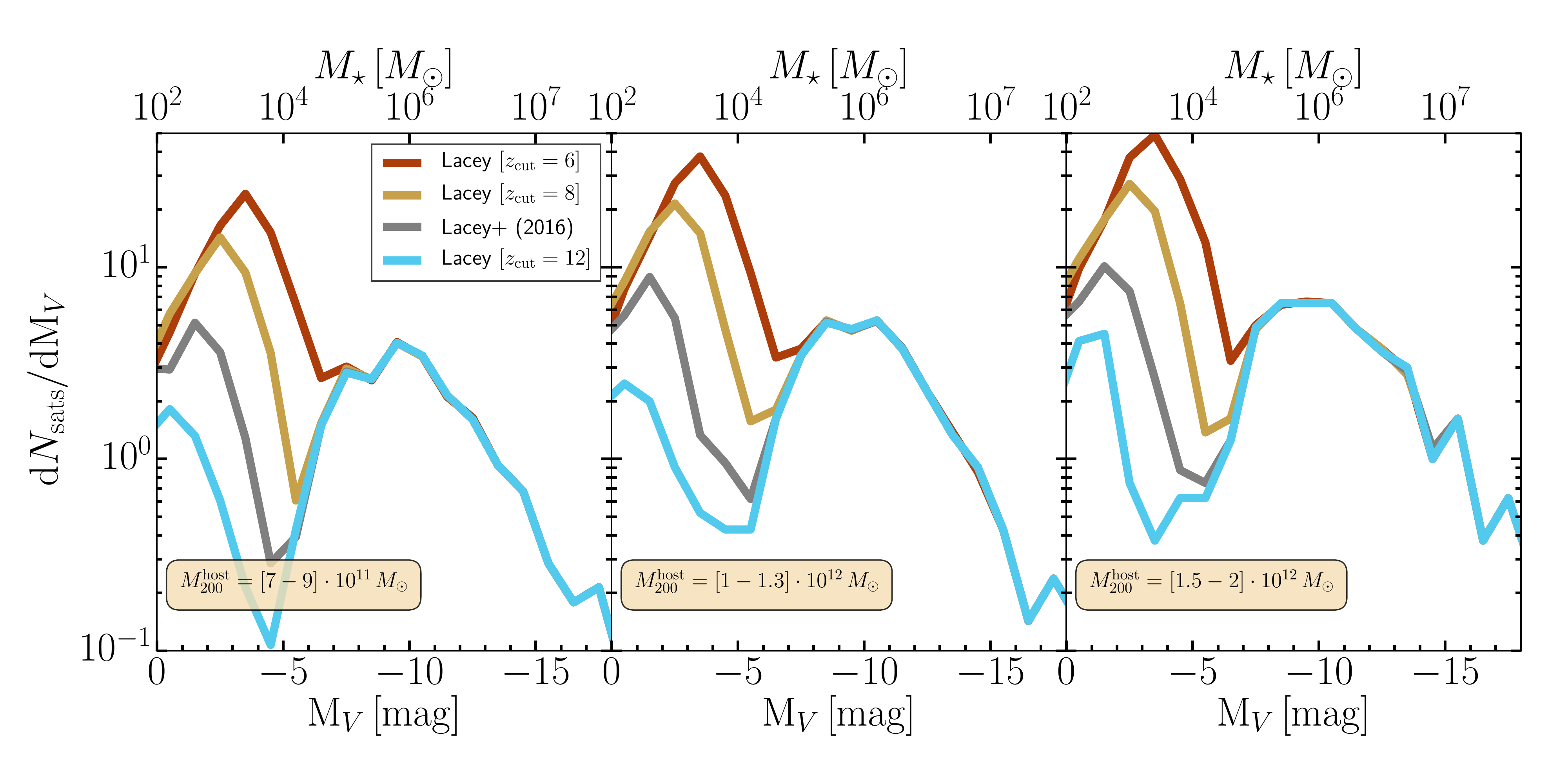}}
  \figcaption{The effect of changing $\zcut$ (i.e., the redshift of
    reionisation) on the differential luminosity function of Milky Way
    satellites predicted by \Galform{}. The fiducial model
    (\citealt{Lacey2016}) assumes $\zcut = 10$ and is shown in
    gray. Variations of $\zcut$ around this value are shown by the
    other curves. Note that $\zcut$ is the only parameter that has
    been varied; in particular, all models assume
    $\vcut = 30\,{\rm kms}^{-1}$.}
\label{fig:zcutVars}
\end{figure*}

\begin{figure}
\center{\includegraphics[width=\columnwidth]{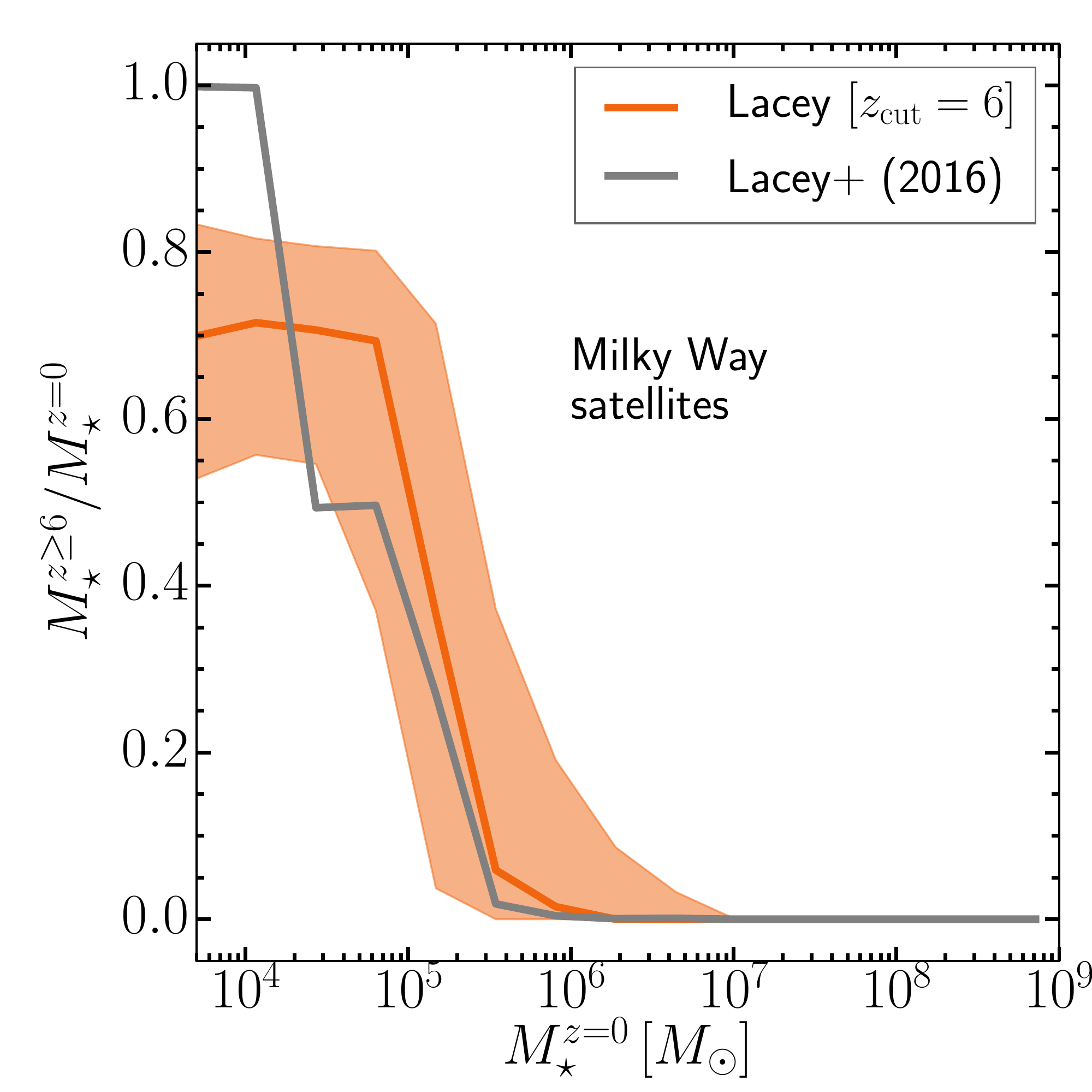}}
\figcaption{Fraction of stellar mass in Milky Way satellites at $z=0$
  that was formed at $z \geq 6$. The solid curves show the mean
  relations averaged over all Milky Way-mass hosts in \coco{};
  the shaded region encompasses 68 per cent of the satellite
  population (shown only for the L16-$z$6 model for clarity).}
\label{fig:reionFrac}
\end{figure}

\begin{figure*}[t!]
\center{\includegraphics[width=\textwidth]{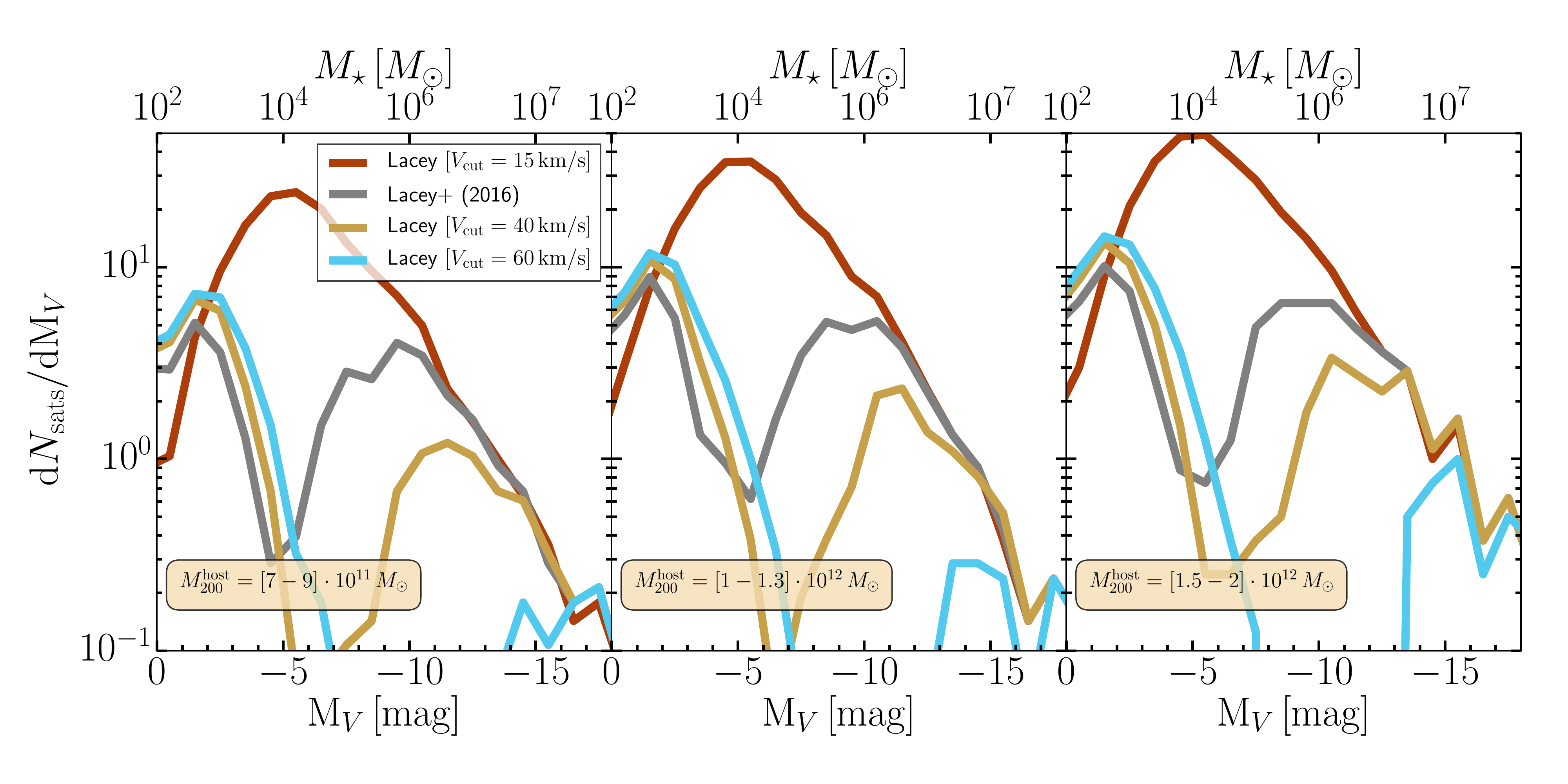}}
\figcaption{As Fig.~\ref{fig:zcutVars}, but now showing the effect of
  changing $\vcut$ (i.e., the threshold halo circular velocity below
  which cooling is suppressed after $z<\zcut$) on the differential
  luminosity function of Milky Way satellites. The fiducial
  \citealt{Lacey2016} model, again shown in gray, assumes $\vcut =
  30\,{\rm kms}^{-1}$. All model variations presented in this figure
  assume $\zcut = 10$.}
\label{fig:VcutVars}
\end{figure*}

\subsubsection{Reionisation in \Galform{}}
\label{sect:reionLF}

Reionisation in \Galform{} is implemented using a simple two parameter
model: to describe the effect of reionisation from a global UV
background, the cooling of gas in a halo with circular velocity, ${\rm
  V}_c$, is turned off if ${\rm V}_c < \vcut$ at $z < \zcut$, where
$\vcut$ and $\zcut$ are input parameters. In this scheme, $\zcut$
controls {\it when} reionisation happens and $\vcut$ determines {\it
  which} haloes are affected by reionisation. While this treatment may
appear oversimplified at first, the $\vcut-\zcut$ approach is in fact
a good approximation to a comprehensive, self-consistent calculation
of reionisation in \Galform{} performed by \cite{Benson2002a}. In
fact, it was shown by \cite{Font2011} that the $\vcut-\zcut$ method
remains a good approximation even when local ionising sources are
included in addition to the global ionising background. This approach
has the added advantage that investigating the effect of changing the
small number of parameters controlling reionisation on the predicted
satellite luminosity function is relatively simple.

We now explore in detail the effects of changing $\zcut$ and $\vcut$
on the shape of the luminosity function of satellites. In what
follows, we will treat the L16 model (in which $\zcut=10$ and
$\vcut=30\,{\rm kms}^{-1}$) as the `fiducial' model against which all
qualitative changes will be compared. All parameters of the galaxy
formation model, apart from $\vcut$ and $\zcut$, are kept fixed at
their L16 values. Throughout this paper, satellites are defined as
\Galform{} galaxies that are located within $r_{200}$ of their host
halo centre. Since we follow the merger trees obtained from the
\coco{} $N$-body simulation, the effects of tidal stripping and
dynamical friction on infalling subhaloes are automatically taken into
account. \Galform{} keeps track of `orphan' galaxies -- those that
have `lost' their subhalo after infall due to limited numerical
resolution. The orbits of these galaxies are followed by tracking the
most bound particle of the subhalo from the last snapshot in which
this particle is associated with a resolved object \cite[][see also
  Appendix C in \citealt{Newton2017}]{Simha2017}.

Fig.~\ref{fig:zcutVars} shows the effect of changing $\zcut$ (the
redshift at which reionisation happens) on the differential luminosity
function. Each of these models assumes $\vcut=30\,{\rm kms}^{-1}$. The
general shape of the curves is similar: the abundance of satellites
slowly increases faintwards of ${\rm M}_V = -16$, peaking at ${\rm
  M}_V = -10$. Fainter than this, all models exhibit a `valley', the
location of which depends only weakly on the value of $\zcut$. The
depth of this valley (and the number of satellites fainter than this)
differs significantly as $\zcut$ varies. In particular, the earlier
the redshift of reionisation, the lower the abundance of galaxies
fainter than ${\rm M}_V=-5$ ($M_\star \approx 10^4\,M_\odot$). This
figure shows that for a fixed value of $\vcut$, the location of the
peaks of the two populations carved out by the reionisation valley is
largely independent of the choice of $\zcut$.

The interpretation of this dependence of the number of faint
satellites on $\zcut$ is straightforward: when reionisation occurs
very early (say at $z=12$), very few haloes with circular velocity
exceeding $\vcut=30\,{\rm kms}^{-1}$ have formed. As a result, cooling
is suppressed in a significant fraction of haloes, preventing the
formation of new satellites (although haloes in which gas has already
cooled can continue to form stars and become brighter). A later
redshift of reionisation allows many more faint galaxies to form
before cooling is prevented. The parameter $\zcut$ therefore affects
the amplitude of the differential luminosity function fainter than the
location of the `valley'. It is also important to note that changing
$\zcut$ has no effect on the bright end of the luminosity function, as
these galaxies primarily assemble much later, long after reionisation
has ended.

The effect of $\zcut$ on the assembly history and abundance of faint
galaxies is demonstrated in Fig.~\ref{fig:reionFrac}, which shows the
fraction of stellar mass assembled before $z=6$ by satellites
identified at $z=0$ in the L16 model ($\zcut=10$), and a variant of
L16 where we have set $\zcut=6$ (to which we refer hereafter as the
L16-$z6$ model). It is evident from this figure that the smallest
galaxies are the ones that form earlier (as expected from the
hierarchical build-up of structure in CDM). The faintest population of
satellites ($M_\star \leq 10^5\,M_\odot$), on average, assembles $\sim
60-70$ per cent of their stellar mass prior to reionisation in the
L16-$z$6 model. By contrast, in the L16 model, in which reionisation
occurs at $z=10$, the faintest satellites, on average, have assembled
100 per cent of their present-day mass by $z=6$. These differences
explain the larger abundance of faint satellites in the L16-$z$6 model
compared to L16 and, by extension, the systematic effect of changing
$\zcut$ on the amplitude of the luminosity function fainter than the
valley.

\begin{figure*}[t!]
\center{\includegraphics[width=\textwidth,trim={0 2.5in 0 0}]{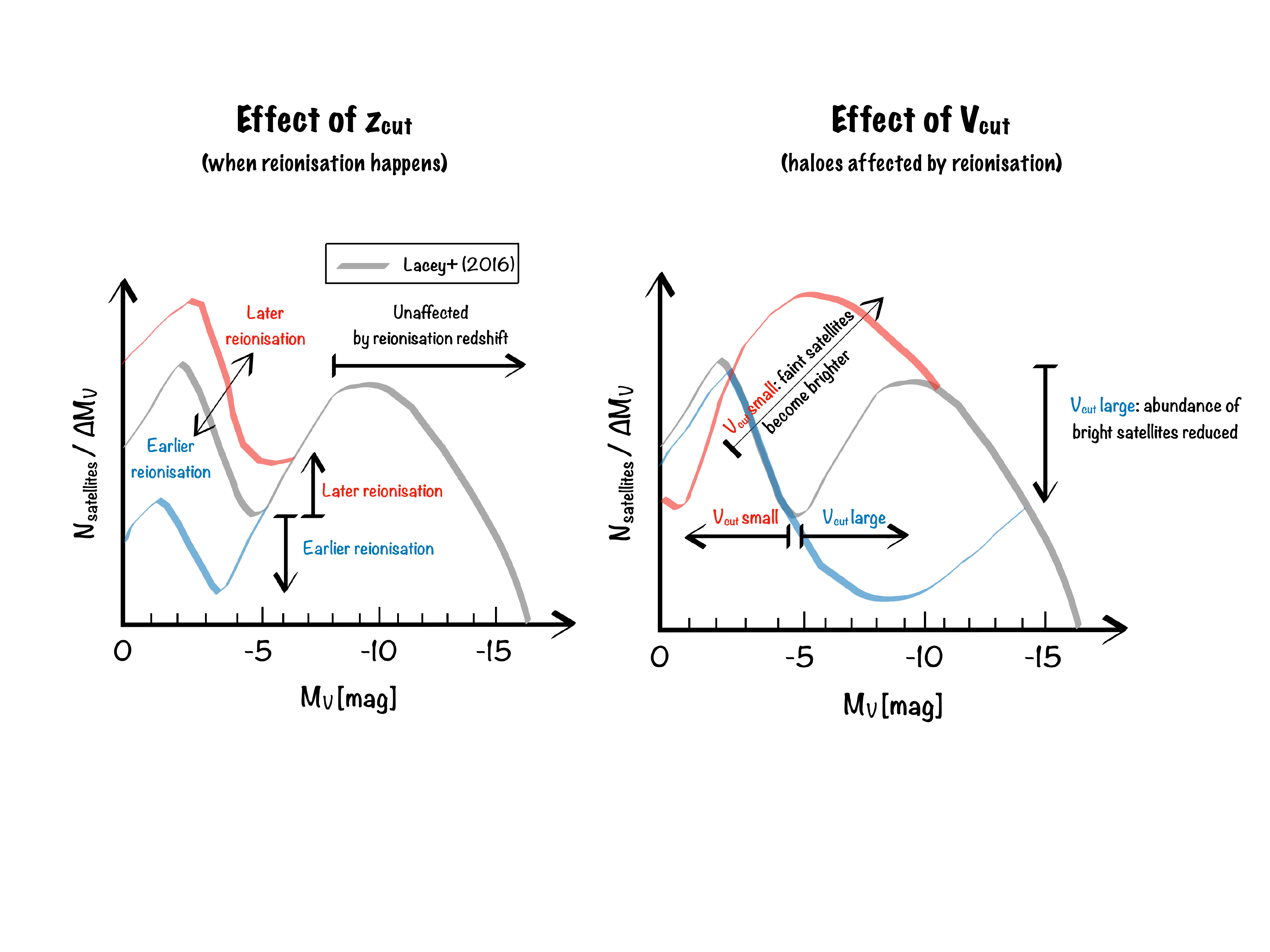}}
\figcaption{A schematic illustration of the role played by $\zcut$ and
  $\vcut$ in shaping the luminosity function of satellite
  galaxies. This figure essentially summarises the behaviours observed
  in Figs.~\ref{fig:zcutVars} and~\ref{fig:VcutVars}. The gray curve
  represents the general shape of the differential luminosity function
  predicted by the \citealt{Lacey2016} model; the red and blue curves
  represent the qualitative response of this base model to changes in
  $\zcut$ and $\vcut$ (left and right panels respectively). }
\label{fig:schematic}
\end{figure*}

The location of the valley itself is instead primarily controlled by
$\vcut$, the threshold that determines which haloes are affected by
reionisation. This is demonstrated in Fig.~\ref{fig:VcutVars}. Here,
we have fixed the value of $\zcut=10$. The L16 model, by default,
assumes $\vcut=30\,{\rm kms}^{-1}$. The effect of changing $\vcut$ is
dramatic across the entire range of luminosities. Increasing $\vcut$
shifts the location of the valley to brighter luminosities. As gas is
prevented from cooling in larger and larger haloes, fewer and fewer
bright galaxies form.

Reducing the value of $\vcut$ to $15\,{\rm kms}^{-1}$, below the
fiducial L16 value (red line in Fig.~\ref{fig:VcutVars}) leaves the
abundance of galaxies brighter than ${\rm M}_V=-10$ unchanged, but
increases the number of faint satellites.  Reionisation now only
affects very small haloes, allowing galaxies in haloes in the range
${\rm V}_c = [15-30]\,{\rm kms}^{-1}$ to grow in stellar mass and
become brighter than in the fiducial case. The bottom of the `valley'
shifts to much fainter magnitudes (${\rm M}_V \approx 0$ in the
example in Fig.~\ref{fig:VcutVars}).

A summary of the numerical experiments performed in this section is
provided in Fig.~\ref{fig:schematic}, which shows a schematic
illustration of the effects of changing $\zcut$ and $\vcut$ on the
shape of the differential luminosity function of satellites. In short,
$\zcut$ (when reionisation takes places) determines the abundance of
satellites fainter than the `valley' in the luminosity function,
leaving the abundance of bright galaxies unaffected. $\vcut$, on the
other hand, determines where exactly the `valley' is formed, and can
influence both the faint and bright ends of the luminosity
function. As the abundance of bright satellites of the Milky Way is
well known, the range of allowed values for $\vcut$ is better
constrained than the value of $\zcut$.

While $\zcut$ and $\vcut$ are input parameters specific to \Galform{},
they are parameterisations of very general properties of the physics
of reionisation: the time when reionisation happens and the mass scale
of haloes that are affected by it. In this sense, the effects
described in this section are generic, and not specific to \Galform{}
or semi-analytic models in general. Indeed, using a formalism similar
to that in \Galform{} to calculate the properties of galactic
subhaloes and a simple prescription for assigning a stellar content to
subhaloes which crudely models the sort of processes that we have
considered here, \cite{Koposov2009} also identified two populations of
satellites. However, their models make rather different predictions to
ours for the properties of the two populations. In
\S\ref{sect:galactic}, we investigate whether the general features
described in this section are present in the observed satellite
luminosity functions of the Milky Way and M31.

\subsubsection{The Lacey et al.  (2016) and Hou et al.  (2016) models of \Galform{}}
\label{sect:specifics}

The most recently published version of \Galform{}, presented in
\cite{Lacey2016}, includes all of the revisions mentioned in
\S\ref{sect:philosophy} in a single, unified model. This model has
been shown to reproduce a wide range of observational relations at
various redshifts such as the fraction of early-type galaxies, the
Tully-Fisher relation, the far-IR number counts, the evolution of the
K-band luminosity function to $z \sim 4$ and the far-UV luminosity
functions at $z \sim 3-10$. In this paper we treat the L16 version of
\Galform{} as the fiducial model against which we compare variations
of \Galform{}. In the published version, L16 assumes $\zcut=10$ and
$\vcut=30\,{\rm kms}^{-1}$. This value of $\vcut$ is consistent with
the hydrodynamical simulations of \cite{Okamoto2008}.

A shortcoming of the L16 model is that the choice of $\zcut=10$ is not
self-consistent. The condition for reionisation may be defined as the
redshift at which $\sim 6$ ionising photons are produced per hydrogen
nucleus\footnote{This threshold ratio can be estimated assuming an
  escape fraction of 20\% and an average of 0.25 recombinations per
  hydrogen atom \citep[see \S2.3 in][]{Hou2016}.}. Counting the total
number of ionising photons produced by galaxies as a function of
redshift in the L16 model, implies that the universe is reionised at
$z \approx 6$, later than the redshift of reionisation inferred from
{\it Planck} \citep[$z=8.8^{+1.7}_{-1.4}$,][]{Planck2016} data. This
discrepancy can be traced back to the strong supernova feedback
implemented in the L16 model, which was required to reproduce the
faint end of the $z=0$ galaxy luminosity function. Strong feedback not
only suppresses the number of ionising photons produced at high
redshift, but also results in the metallicities of Milky Way
satellites being too low when compared with observations \citep[][H16
  hereafter]{Hou2016}.

To remedy these problems, H16 proposed a feedback prescription in
which the feedback strength varies not only as a function of halo
circular velocity (as in all models of \Galform{}) but also with
redshift. In particular, feedback becomes weaker at high redshift
($z>4$), resulting in the production of more ionising photons, and
therefore an earlier redshift of reionisation. The redshift dependence
was chosen to emulate the dynamical supernova feedback model of
\cite{Lagos2013}, which attempts to capture the relationship between
the efficiency of feedback and properties of the ISM, including gas
density, metallicity and molecular gas fractions. Reionisation in the
H16 model occurs at $z=\zcut=7.9$; it also sets $\vcut=30\,{\rm
  kms}^{-1}$. This model also produces an acceptable luminosity
function (${\rm M}_V \lesssim -4$) and metallicity-luminosity relation
for Milky Way satellites.  We will present the detailed predictions of
the L16 and H16 models of \Galform{} in \S\ref{sect:lumMW}.

\section{The combined Milky Way-M31 satellite luminosity function}
\label{sect:galactic}

\begin{figure}[t!]
  \center{\includegraphics[width=\columnwidth]{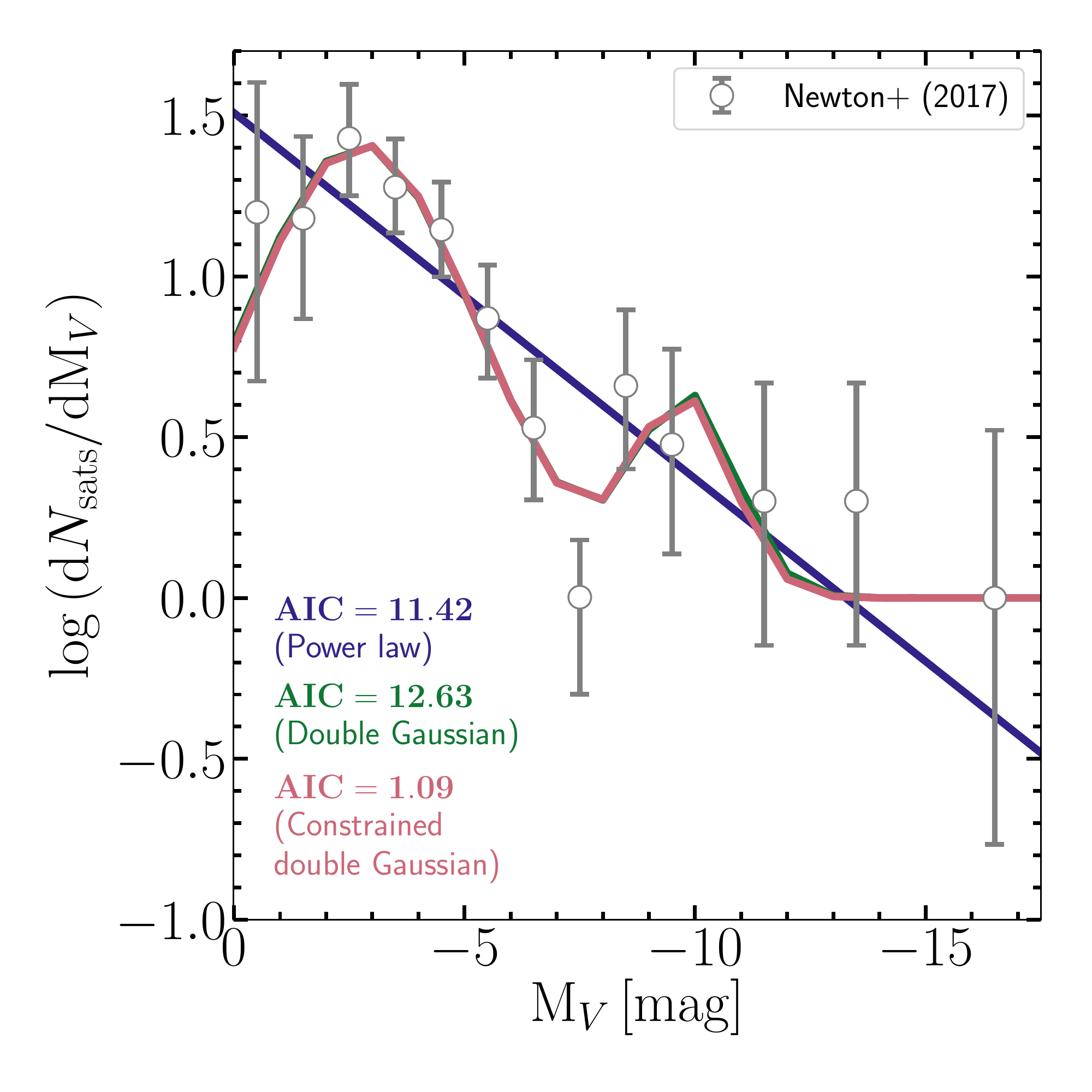}}
  \figcaption{Comparison of fits to the Milky Way satellite luminosity
    function for three models: a power law, a double Gaussian and a
    constrained double Gaussian (see text for details of the models).
    The gray data points show the luminosity function estimated by
    \citealt{Newton2017}. For the classical satellites, error bars
    show Poisson errors; for fainter satellites, they show the
    $1\sigma$ uncertainty estimated by \citealt{Newton2017}. The
    Akaike Information Criterion (AIC) value for each model is given
    in the bottom left.}
\label{fig:AICtest_noM31}
\end{figure}

The (relatively) large number of satellites now known to orbit around
the Milky Way and M31 invites us to investigate if the features in the
luminosity function of satellite galaxies that our models predict (see
\S\ref{sect:reionLF}) are present in the data. 

A total of 54 satellites around the Milky Way have now been detected
in the SDSS and DES. This census is incomplete because the combined
sky coverage of SDSS and DES is only $\sim$ 47 per cent, and both
surveys are subject to detection limits that depend on the satellite
luminosity and distance.

Extrapolations based on $N$-body simulations
\citep{Koposov2008,Tollerud2008,Hargis2014} have suggested that the
total population count is at least $70^{+65}_{-30}$ (at 98 per cent
confidence) for satellites brighter than ${\rm M}_V = -2.7$. Recently,
\cite{Newton2017} applied a new Bayesian method to a sample that
includes the newly detected satellites in SDSS DR9 and DES. They
estimate a total of $124^{+40}_{-28}$ (at 68 per cent confidence)
satellites brighter than ${\rm M}_V = 0$. The Newton et al. estimate
is particularly important for our test because the faint end of the
luminosity function is especially sensitive to the physics of
reionisation.

\begin{figure*}[t!]
  \center{\includegraphics[scale=0.65]{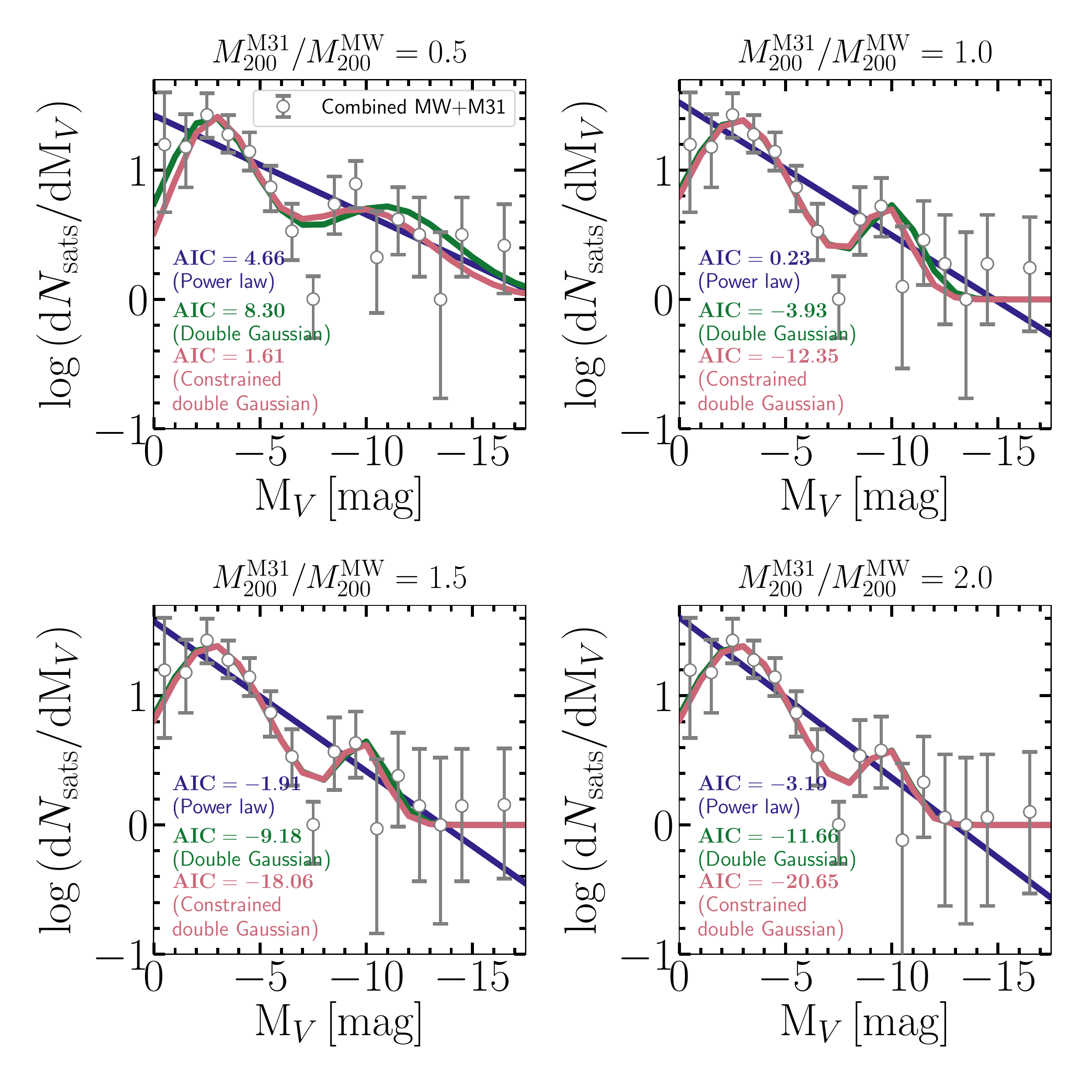}}
  \figcaption{Same as Fig.~\ref{fig:AICtest_noM31}, but for the the
    combined Milky Way-M31 satellite luminosity function (see
    \S\ref{sect:galactic} for details). The different panels
    correspond to different values of $M_{200}^{{\rm M31}} /
    M_{200}^{{\rm MW}}$, as indicated in the labels. The AIC for each
    model is given in each panel. In this figure we have fixed the
    value of $M_{200}^{{\rm MW}} = 10^{12} M_\odot$. AIC statistics
    for other choices of $M_{200}^{{\rm MW}}$ are listed in
    Table~\ref{tab:AICvariants}.}
\label{fig:AICtest}
\end{figure*}

To test the predictions of \Galform{} against this dataset, we assume
that the luminosity function of the pre- and post-reionisation
satellite populations can each be approximated by a Gaussian.  Each
Gaussian has three free parameters controlling the location of the
peak, the width (standard deviation) and height; in total, a general
double Gaussian model has six free parameters. However, our {\em a
  priori} standard theoretical model, which has $\vcut=30 \, {\rm
  kms}^{-1}$, has two fewer degrees of freedom because the locations
of the two peaks (at ${\rm M}_V \approx -10$ and ${\rm M}_V \approx
-3$) are predicted by the model (approximately independently of halo
mass and the value of $\zcut$; see Fig.~\ref{fig:zcutVars}). We refer
to this as the {\it constrained} double Gaussian model. Finally, we
compare the double Gaussian models to the simplest possible model of
the satellite luminosity function, a power law.

\begin{table*}
  \centering
  \caption{Summary of AIC statistics comparing the power law (PL),
    double Gaussian (DG) and constrained double Gaussian (CDG) models
    of the combined Milky Way and M31 satellite luminosity
    functions. Each column corresponds to a different choice for the
    mass of the M31 halo relative to that of the Milky Way halo; each
    row corresponds to a different choice for the mass of the Milky
    Way halo. The preferred model in each case (highlighted in {\bf
      bold}) is the model with the lowest AIC value.}

\begin{tabular}{c|cccc}
\hline \hline
$M_{200}^{{\rm MW}}$ & \multicolumn{4}{c}{Ratio [$M_{200}^{{\rm M31}}/M_{200}^{{\rm MW}}$]} \\
$[M_\odot]$                   & 0.5 & 1.0 & 1.5 & 2.0 \\
\hline
$5 \times 10^{11}$ & PL  = 4.15  & PL = -1.05  & PL = -3.07 & PL = -4.23 \\
                   & DG  = 7.38   & DG = -7.23  & DG = -11.45 & DG = -16.87   \\
                   & {\bf CDG = 0.30}  & {\bf CDG = -15.97} & {\bf CDG = -20.43} & {\bf CDG = -22.26} \\
\hline
$1 \times 10^{12}$ & PL  = 4.66  & PL = 0.23  & PL = -1.91 & PL = -3.19 \\
                   & DG  = 8.30   & DG = -3.93  & DG = -9.18 & DG = -11.66   \\
                   & {\bf CDG = 1.61}  & {\bf CDG = -12.35} & {\bf CDG = -18.06} & {\bf CDG = -20.65} \\
\hline
$1.5 \times 10^{12}$ & PL  = 5.57  & PL = 1.11  & PL = -1.09 & PL = -2.44 \\
                   & DG  = 9.83   & DG = -1.47  & DG = -7.33 & DG = -10.28   \\
                   & {\bf CDG = 2.98}  & {\bf CDG = -9.64} & {\bf CDG = -16.07} & {\bf CDG = -19.22} \\
\hline
$2 \times 10^{12}$ & PL  = 6.27  & PL = 1.79  & PL = -0.45 & PL = -1.84 \\
                   & DG  = 10.98   & DG = 0.52  & DG = -5.73 & DG = -9.04   \\
                   & {\bf CDG = 4.10}  & {\bf CDG = -7.44} & {\bf CDG = -14.33} & {\bf CDG = -17.91} \\
\hline \hline
\end{tabular}
\label{tab:AICvariants}
\end{table*}

To determine which model is preferred by the data, we apply the Akaike
Information Criterion \citep[AIC;][]{Akaike1974}.  The AIC gives a
measure of the relative quality of different models given the data and
is therefore very useful for model selection. It penalises models with
a larger number of parameters; the model with the lowest AIC value is
preferred.  For two models, $A$ and $B$, with corresponding AIC values
AIC$_A$ and AIC$_B$, the quantity $\exp\left[ \frac{1}{2} ({\rm AIC}_A
  - {\rm AIC}_B) \right]$ may be interpreted as the relative
likelihood of model $A$ over model $B$.  For our analysis we consider
a variant of the AIC that corrects for small sample size
\citep{Burnham2003}.

Fig.~\ref{fig:AICtest_noM31} shows fits of our three models to the
Milky Way satellite luminosity function and the associated AIC values.
While the goodness-of-fit is best for the general double Gaussian, the
AIC penalises that model for having six free parameters (compared to
only two for the power law). The constrained double Gaussian, where
two parameters are fixed according to predictions of our galaxy
formation model, strongly improves the AIC value. In the Milky Way
data alone, therefore, there is evidence for the presence of a bimodal
population of satellites just as our model predicts.

To maximise the statistical power of the test, we combine the
satellite populations of the Milky Way and M31 using the strategy that
we now describe. We combine the satellite luminosity function for the
Milky Way estimated by Newton et al. with the satellite luminosity
function for M31 compiled from the Pan-Andromeda Archaeological Survey
\citep[PAndAS;][]{McConnachie2009,McConnachie2012,Ibata2014,Martin2016}.
PAndAS has surveyed the region within $\sim150$ kpc (in projection) of
the centre of M31, but the census of satellites is by no means
complete. Limiting the sample to satellites brighter than ${\rm M}_V =
-9.13$, \cite{Ferrarese2016} find a total of 19 satellites in
M31. Here, we make the assumption that the PAndAS sample of satellites
is complete to about ${\rm M}_V = -8$ \citep[c.f.][]{McConnachie2012},
which extends the sample of M31 satellites to 23.

To combine the satellite luminosity functions of the Milky Way and M31
we assume that the abundance of satellites scales with the mass of the
central galaxy's halo \citep{Wang2012}.  We consider four values for
the ratio of the masses of the two haloes: $M_{200}^{{\rm M31}} /
M_{200}^{{\rm MW}} = [0.5,1.0,1.5,2.0]$.  For a given value of
$M_{200}^{{\rm MW}}$ we can then derive a corresponding value of
$r_{200}$ for M31 and extrapolate the PAndAS luminosity function to
$r_{200}$ by assuming a radial profile for the distribution of
satellites. Following Newton at al., we assume that the radial number
density of satellites, $n(r)$, follows an \cite{Einasto1965} profile:
\bq
\frac{n(\chi)}{\left<n\right>} = \frac{\alpha c_{200}^{\;3}}{3 \left( \frac{\alpha}{2} \right)^{\frac{3}{\alpha}} \gamma \left( \frac{3}{\alpha}, \frac{2}{\alpha} c_{200}^{\; \alpha} \right)} \exp \left[ -\frac{2}{\alpha} \left( c_{200} \chi \right)^\alpha \right],
\eq
where $\chi=r/r_{200}$, $\left<n\right>$ is the mean number density
within $r_{200}$, $c_{200}=4.9$, $\alpha=0.24$ and $\gamma$ is the
lower incomplete Gamma function.

For satellite magnitudes sampled in both galaxies and for a given
value of $M_{200}^{{\rm MW}}$, the combined luminosity function is the
average of the Milky Way and M31 estimates extrapolated to $r_{200}$,
with the latter rescaled by the ratio of the M31 and Milky Way halo
masses. As the estimate for M31 satellites is limited to galaxies
brighter than ${\rm M}_V = -8$, only the Newton et al. estimate for
the Milky contributes to bins fainter than this.

Fig.~\ref{fig:AICtest} shows fits of our three models to the combined
Milky Way and M31 satellite luminosity function for different values
of $M_{200}^{{\rm M31}} / M_{200}^{{\rm MW}}$. Here, we have fixed
$M_{200}^{{\rm MW}} = 10^{12} M_\odot$. In each panel we list the
corresponding AIC values for each of the best fitting power law,
double Gaussian and constrained double Gaussian models. It is clear
that the constrained double Gaussian model is preferred by the
data. As we have already seen from Fig.~\ref{fig:AICtest_noM31}, the
smaller number of free parameters in this model results in a
significant improvement in its AIC value over the general double
Gaussian.

The results of this test for other choices of $M_{200}^{{\rm MW}}$ are
summarised in Table~\ref{tab:AICvariants}.  In {\it all} cases, the
presence of two populations in the observed luminosity function is
significantly preferred over a single power law. The value of the AIC
for the power law varies very little with either the mass of the Milky
Way halo or the ratio of the masses of the M31 and Milky Way
haloes. The values for the double Gaussian and constrained double
Gaussian, on the other hand, decrease significantly with
$M_{200}^{{\rm M31}}/M_{200}^{{\rm MW}}$ and the difference relative
to the power law is largest when the M31 halo is assumed to be twice
as massive as the Milky Way halo.

\begin{figure*}[t!]
  \center
      {\includegraphics[width=\textwidth,angle=0]{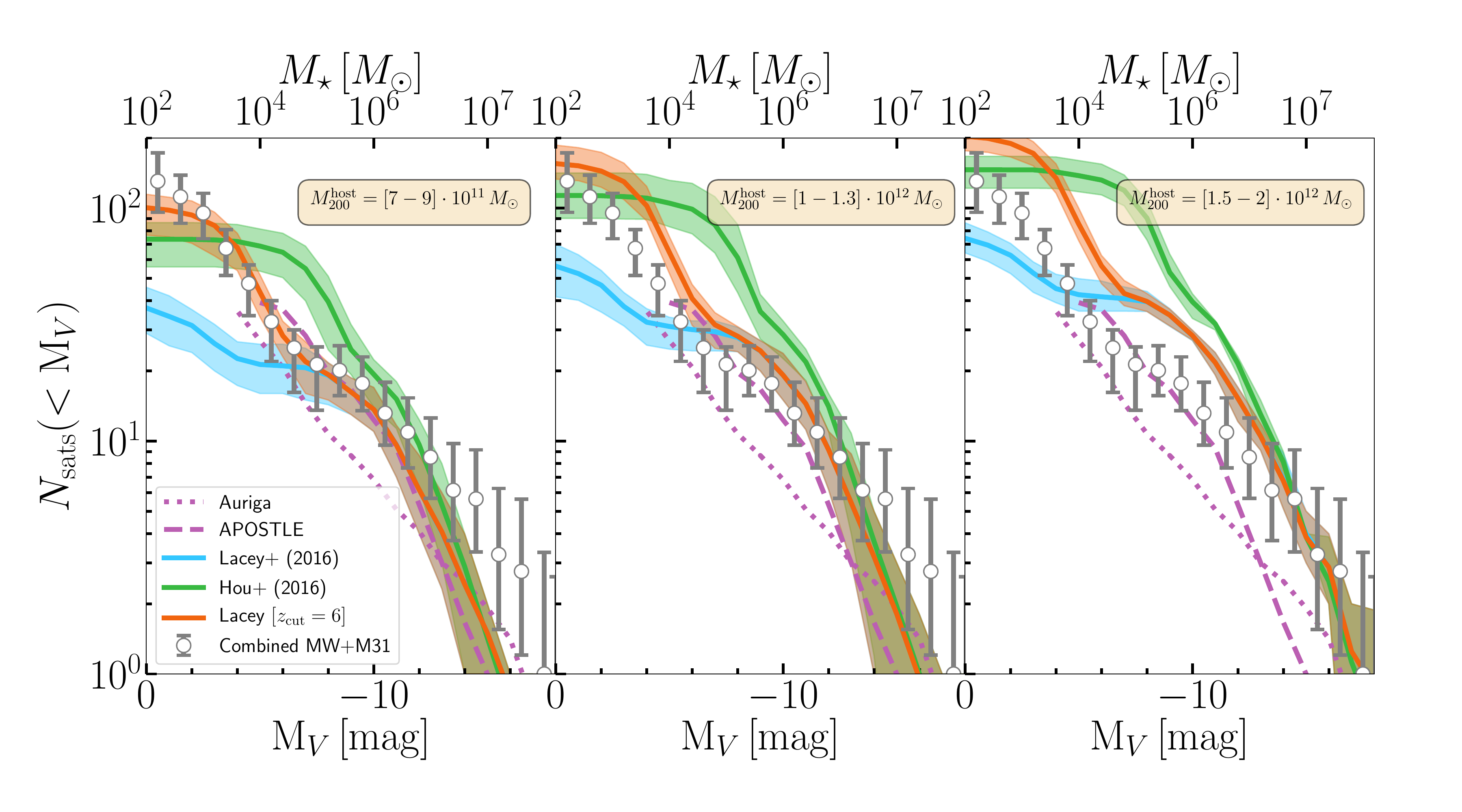}}
      \figcaption{The average cumulative satellite luminosity function
        of the Milky Way and M31, as a function of absolute V-band
        magnitude, ${\rm M}_V$ (lower axis) and present-day stellar
        mass, $M_\star$ (upper axis). \Galform{} uses stellar
        population synthesis models to convert stellar SEDs into
        broad-band luminosities and magnitudes. Each panel presents
        the satellite luminosity function predicted by \Galform{}
        measured in different bins of host halo mass, $M_{200}$
        (assumed to be the same for the Milky Way and M31). The
        coloured solid lines show the mean prediction of various
        \Galform{} models as described in the main text; the
        associated shaded regions mark the 10$^{\rm th}$ and 90$^{\rm
          th}$ percentile spread around the mean relation. The dashed
        and dotted magenta curves, respectively, show the results from
        the \Apostle{} and \Auriga{} hydrodynamical simulations; these
        curves are for $\sim 10^{12} M_\odot$ haloes and are truncated
        below magnitudes (stellar masses) at which resolution effects
        become important. Finally, the gray points represent the
        combined Milky Way+M31 satellite luminosity function obtained
        as described in \S\ref{sect:galactic} with Poisson errors for
        satellites brighter than ${\rm M}_V=-8$ and the 1$\sigma$
        uncertainty estimated by Newton et al. for satellites fainter
        than this.}
\label{fig:cumulativeAll}
\end{figure*}

\begin{figure*}
  \center{\includegraphics[width=\textwidth]{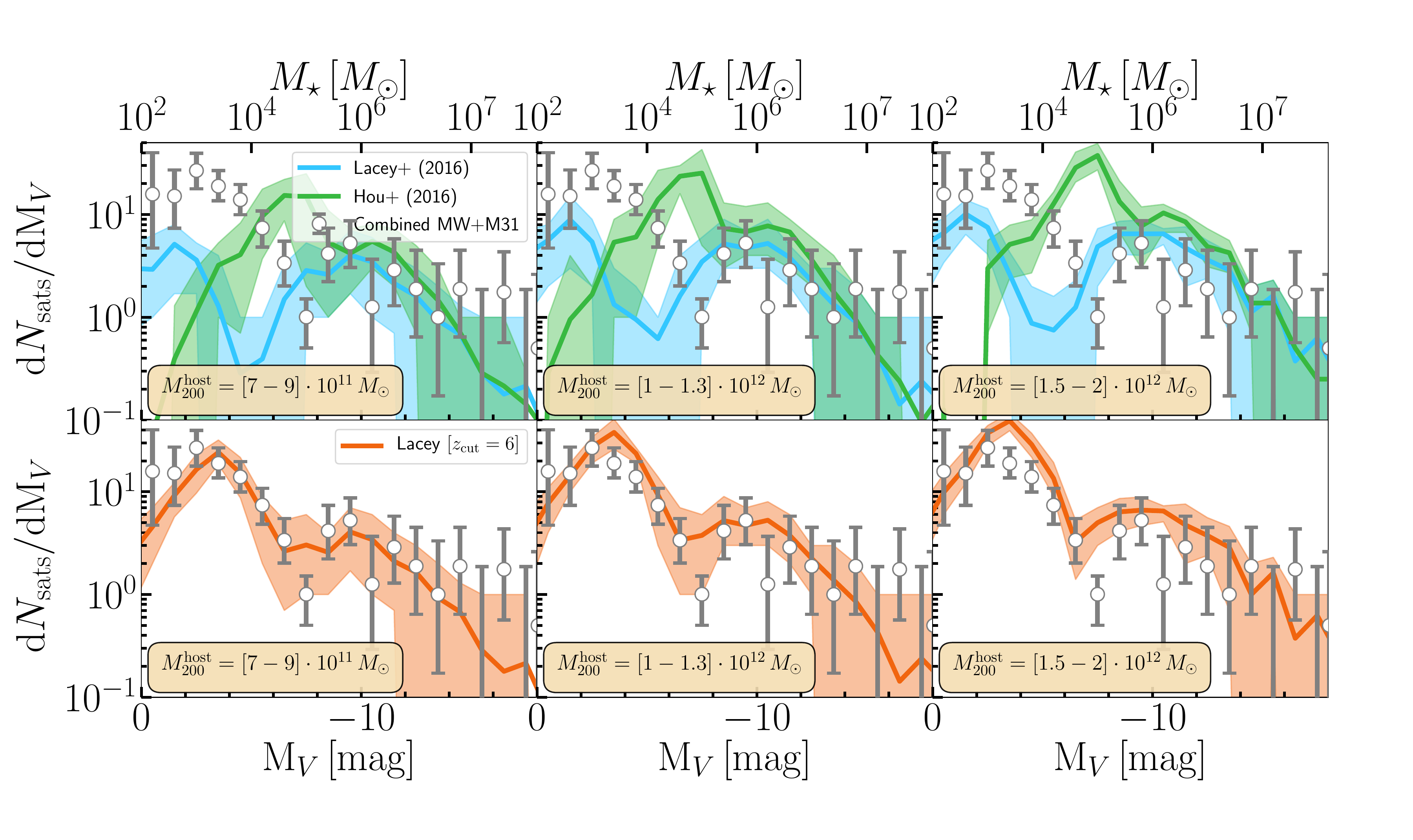}}
  \figcaption{As Fig.~\ref{fig:cumulativeAll}, but now showing the
    average of the differential luminosity function of satellites in
    the Milky Way and M31.}
\label{fig:differentialAll}
\end{figure*}

\section{Comparison of models with the data and predictions}
\label{sect:lumMW}

\subsection{The luminosity function of the Milky Way-M31 satellites}

We have seen in \S\ref{sect:galactic}, that both the Milky Way and the
combined Milky Way-M31 satellite luminosity functions are best
described as the sum of two distinct populations, each characterised
by a Gaussian.  We now consider how the predictions of specific
\Galform{} models compare with the data. For definitiveness, we will
take $M_{200}^{{\rm M31}}/M_{200}^{{\rm MW}}=1$, but our conclusions
are not affected by this choice: the inclusion of the M31 satellites
affects only the bright end of the satellite luminosity function not
the faint population that includes most of the satellites.

Fig.~\ref{fig:cumulativeAll} compares the cumulative satellite
luminosity function for a variety of \Galform{} models with the
combined Milky Way-M31 estimate (gray data points). In addition to the
L16 and H16 models described in \S\ref{sect:specifics}, we have also
included the L16-$z$6 model, which, we recall, is identical to L16
except that $\zcut=6$, which is the self-consistent value for the
redshift of reionisation in the L16 model. It should be noted that
according to our approximate method to determine when reionisation
actually occurs in the model (based on counting the total number of
ionising photons produced per hydrogen nucleus), the L16-$z$6 model
does, indeed, reionise the Universe at $z=6$, as in the default L16
model. The reason for this is that changing the value of $\zcut$ only
affects the abundance of galaxies with $M_\star \lesssim 10^5 M_\odot$
(see \S\ref{sect:reionLF} and Fig.~\ref{fig:reionFrac}), far below the
scale of the dominant sources of ionising photons at $z \geq 6$
\citep[typically $M_\star \geq 10^7
  M_\odot$,][]{Hou2016,Sharma2016}. Therefore, reasonable changes to
the value of $\zcut$ do not affect the time at which the model
satisfies the condition for reionisation.

Each panel in Fig.~\ref{fig:cumulativeAll} shows the luminosity
function for different ranges of the assumed mass of the Milky Way
host halo; the shaded regions mark the 10$^{{\rm th}}$-90$^{{\rm th}}$
percentile spread in the predicted number counts for that mass
bin. The dotted and dashed curves in magenta, respectively, represent
the mean predictions from the \Auriga{} \citep{Simpson2017} and
\Apostle{} \citep{Fattahi2016,Sawala2016} hydrodynamical simulations
in which the host halo mass is $\sim 10^{12} M_\odot$. These curves
are only plotted down to stellar masses where the simulations are well
resolved. Neither \Apostle{} nor \Auriga{} have sufficient resolution
to follow the luminosity function to fainter magnitudes.

It is interesting to note that all three variants of \Galform{} seem
to prefer relatively low masses for the Milky Way (and M31) halo
($M_{200} = [0.7-1.4] \times 10^{12} \, M_\odot$). While the L16, H16
and L16-$z$6 models roughly match the cumulative number counts for
satellites brighter than ${\rm M}_V = -10$, significant differences
can be seen at fainter magnitudes.

The L16 model, for example, is in good agreement with the data down to
${\rm M}_V \sim -5$ ($M_\star \sim 10^4 \, M_\odot$), but
underpredicts the number of satellites fainter than this
magnitude. While this may at first appear to be a consequence of the
strong feedback employed in the L16 model, it is, in fact, a result of
the choice of $\zcut=10$. This can be seen by comparing the prediction
of L16 to L16-$z$6, which agrees very well with the data down to the
faintest magnitudes. In this model, the strength of supernova feedback
as a function of halo mass is identical to that in L16; the only
difference is that reionisation now occurs later, at $z=6$, rather
than at $z=10$. Since reionisation is delayed, gas can now cool in
haloes with ${\rm V}_c < \vcut = 30\,{\rm kms}^{-1}$ for a longer
period of time, allowing the abundance of the faint galaxy population
to build up.

While the total number of satellites brighter than ${\rm M}_V=0$
predicted by the H16 model is consistent (within $\sim 2\sigma$) with
the observed total number, the overall shape of the predicted
luminosity function is not consistent with the data. This difference
is clearly seen in Fig.~\ref{fig:differentialAll}, which shows the
differential satellite luminosity function. This figure shows that the
H16 model overpredicts the abundance of galaxies in the range $-9 \leq
{\rm M}_V \leq -5$ and vastly underpredicts it at fainter
magnitudes. This behaviour can be attributed to the weaker feedback at
high redshift in H16, which allows faint galaxies to build up their
stellar mass, shifting their occupancy from fainter to brighter
magnitudes in the luminosity function. This results in a shape that is
inconsistent with the Milky Way-M31 data. On this evidence, the H16
model is ruled out by the Milky Way data. A similar case can be made
for the default L16 model, which, as shown in
Fig.~\ref{fig:differentialAll}, greatly underpredicts the abundance of
satellites fainter than ${\rm M}_V = -7$.

As we have already seen in Fig.~\ref{fig:cumulativeAll}, the L16-$z$6
model provides an excellent match to the data for the Milky Way. This
is demonstrated in greater detail in the lower panels of
Fig.~\ref{fig:differentialAll}, where we can see that even the shape
of the observed differential luminosity function is captured almost
perfectly by the L16-$z$6 model, particularly for the lowest mass
bins. We recall that $\zcut=6$ was not chosen to provide a good match
to the luminosity function; it is the self-consistent value for the
redshift of reionisation appropriate to the L16 model. Although the
remarkable level of agreement between this model and the data may well
be fortuitous given the noisy data, it is interesting, nevertheless,
that the main features of the shape of the observed luminosity
function are reproduced by the model. The discovery of new,
ultra-faint satellites will help to confirm or exclude the
predicitions of this model. Importantly, other predictions of the
L16-$z$6 model (such as the field luminosity functions, Tully-Fisher
relation, etc.)  that the L16 model reproduces are unaffected by this
change in $\zcut$. This is because the bright galaxies that these
observations probe are not sensitive to {\it when} exactly
reionisation happens, as we demonstrated in
\S\ref{sect:reionLF}. Finally, it is worth noting that, as
Fig.~\ref{fig:differentialAll} shows, the general shape of the
differential luminosity function predicted by a given \Galform{} model
is independent of the host halo mass.

\begin{figure*}
\center{\includegraphics[width=\textwidth]{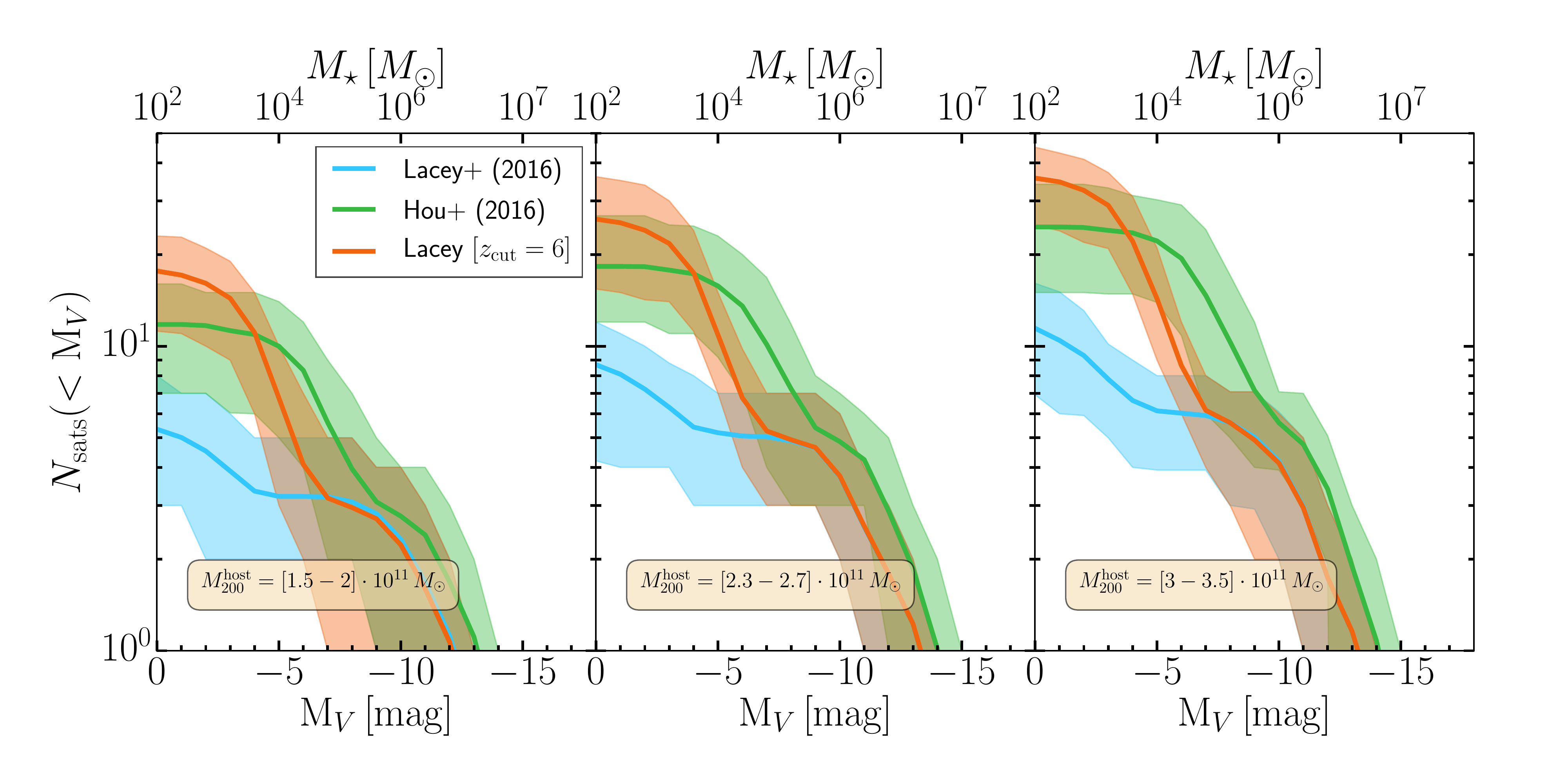}}
\figcaption{The cumulative luminosity function of satellites in
  LMC-mass hosts predicted by various \Galform{} models. The total
  number of predicted satellites is strongly correlated with the
  assumed mass of the LMC-mass dark matter halo, as shown in the
  different panels. All models assume $\vcut = 30\,{\rm kms}^{-1}$.}
\label{fig:cumulativeLMC}
\end{figure*}

\begin{figure*}
\center{\includegraphics[width=\textwidth]{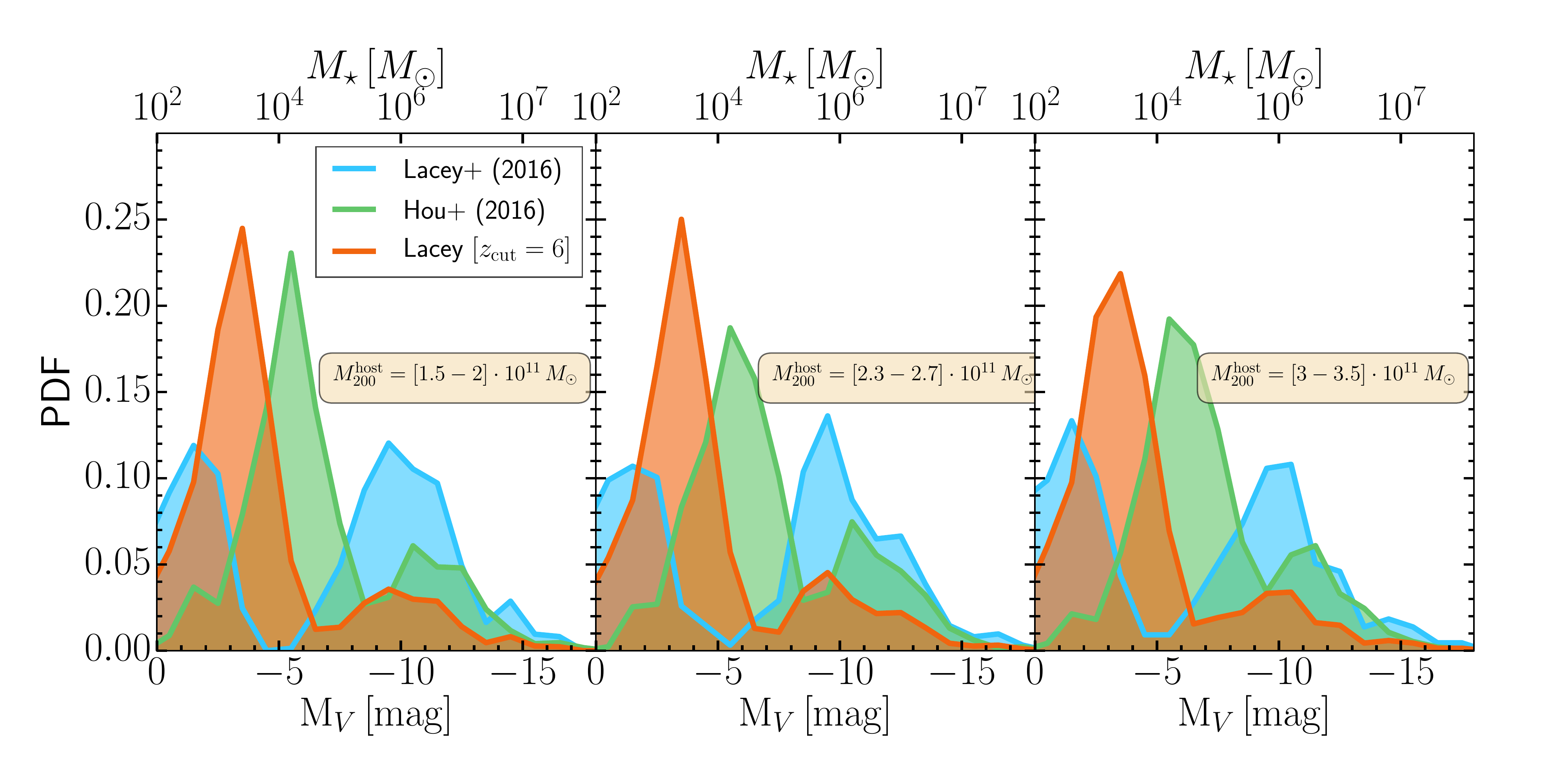}}
\figcaption{Probability distribution functions of satellites in
  LMC-mass hosts predicted by three different \Galform{} models.}
\label{fig:pdfLMC}
\end{figure*}

\subsection{Predictions for the luminosity function of LMCs}
\label{sect:predLMC}

To conclude this section, we provide predictions for the luminosity
function of satellites of LMC-mass galaxies, in anticipation of future
surveys like the LSST \citep{Ivezic2008} and WFIRST
\citep{Spergel2015} that may detect satellites of such systems. The
luminosity functions of lower mass hosts are especially interesting as
the abundance of their faint satellites is particularly sensitive to
the details of reionisation.

Fig.~\ref{fig:cumulativeLMC} presents the cumulative satellite
luminosity function of LMC-mass systems predicted by the L16, H16 and
L16-$z$6 models of \Galform{}. As in Fig.~\ref{fig:cumulativeAll},
each panel shows the luminosity function in different bins of mass for
the LMC-mass host. The systematic difference between the three models
is consistent with the trends seen for Milky Way-mass hosts, with the
L16-$z$6 model predicting the largest number of satellites brighter
than ${\rm M}_V=0$ in each mass bin. For example, the L16 model, on
average, predicts fewer satellites in the highest bin of host halo
mass (12) than the L16-$z$6 model does in the lowest mass bin
(18). The halo mass for the LMC itself is close to $\sim 2.5 \times
10^{11} \, M_\odot$ \citep[e.g.][]{Penarrubia2016}, for which the
L16-$z$6 model predicts $26 \pm 10$ (68 per cent confidence)
satellites brighter than ${\rm M}_V=0$.

Differences in the predictions of the \Galform{} models are revealed
explicitly in Fig.~\ref{fig:pdfLMC}, which shows the PDF of the LMC
satellite luminosity function (i.e., the probability that a satellite
occupies a particular magnitude bin). This figure is simply the
differential luminosity function of LMC satellites normalised by the
total number of satellites. The PDFs in each of the L16, H16 and
L16-$z$6 models are qualitatively similar, exhibiting a bimodal
population in all cases. All three models peak at ${\rm M}_V \approx
-10$, before displaying the characteristic `reionisation valley' we
have previously seen in \S\ref{sect:reionLF}. The distributions then
peak once more at magnitudes fainter than the location of this
valley. Whereas the fraction of galaxies in both peaks is comparable
in the L16 model, the L16-$z$6 model predicts $\sim 10$ times as many
faint satellites (${\rm M}_V \geq -5$) as bright ones.

Recently, \cite{Dooley2017} made use of abundance matching to infer
the total satellite population around LMC-mass hosts. As shown by
these authors, the different abundance matching models available in
the literature predict very different numbers of satellites,
particularly at the faint end of the luminosity function. In their
work, one of the abundance matching models tested is the one
calibrated by \cite{GarrisonKimmel2017}, and predicts $\sim 16$
satellites more massive than $10^3 M_\odot$, comparable to what is
predicted by the L16-$z$6 model ($\sim 18$) for the lowest LMC mass
bin. However, \cite{Sawala2015} have shown, using the \Apostle{}
hydrodynamical simulations of the Local Group, that standard abundance
matching prescriptions such as those on which these numbers are based
are invalid for galaxies with stellar mass less than $\sim 10^6
M_\odot$ or halo mass less than $\sim 3 \times 10^8 M_\odot$ because
only a decreasing fraction of haloes below this mass host a visible
galaxy.

\section{Conclusions}
\label{sect:Conclusions}

The luminosity function of dwarf satellites is one of the most
informative statistics of the galaxy population that can be measured
from local observations. The total number of satellites is sensitive
to the physics of reionisation, the strength of supernova feedback,
the host halo mass and the nature of the dark matter. In this paper,
we have explored the way in which reionisation influences both the
amplitude and the shape of the satellite luminosity function.

To obtain a well-resolved sample of Milky Way and LMC-mass haloes, we
made use of the high-resolution {\it Copernicus Complexio} (\coco{})
suite of simulations \citep{Hellwing2016,Bose2016}. The merger trees
of the dark matter haloes in \coco{} were populated with galaxies
using the Durham semi-analytic model of galaxy formation, \Galform{}
\citep{Cole2000}. \Galform{} is a flexible tool that allows the
parameter space of galaxy formation models to be explored
efficiently. In this model, reionisation is characterised by two
parameters: $\zcut$, which determines the redshift at which
reionisation is complete, and $\vcut$, which controls the mass scale
of haloes that are affected by reionisation. To emulate the net effect
of an ionising background, gas cooling in haloes with circular
velocity, ${\rm V}_c < \vcut$, is suppressed at redshifts
$z<\zcut$. \cite{Benson2002a} showed that this simple prescription
agrees remarkably well with the results of a detailed, self-consistent
model for the coupled evolution of the global properties of the
intergalactic medium and the formation of galaxies in the presence of
a photoionising background due to stars and quasars. Thus, while
$\zcut$ and $\vcut$ are parameters specific to \Galform{}, they
quantify general features of the effects of reionisation on galaxy
formation.

In this paper we have considered two recent versions of \Galform{}:
the fiducial \cite{Lacey2016} (L16) model which assumes $\zcut=10$ and
$\vcut=30\,{\rm kms}^{-1}$, and the \cite{Hou2016} (H16) model, which
assumes $\zcut=7.9$ and $\vcut=30\,{\rm kms}^{-1}$. The two models
differ only in their treatment of supernovae feedback: whereas the
strength of feedback in L16 is a function of halo circular velocity
only, supernovae feedback in H16 varies as a function of circular
velocity and redshift, becoming weaker at $z>4$ for the reasons
explained in \S\ref{sect:specifics}. To understand the effects of
reionisation on the satellite luminosity function, we additionally
considered departures from the L16 model, varying $\vcut$ and $\zcut$
about their fiducial values.  Fig.~\ref{fig:schematic} illustrates the
effects of varying $\zcut$ and $\vcut$ on the amplitude of the faint
end and the overall shape of the satellite luminosity function. The
general picture that emerges is:
\begin{enumerate}
\item The general shape of the differential satellite luminosity
  function, exhibits two peaks: one corresponding to a population of
  faint galaxies that were mostly assembled before reionisation and
  one corresponding to a population of bright galaxies that were
  mostly assembled after reionisation.  These features are generic and
  do not depend on the details of the \Galform{} model.

\item Between these peaks there is a `valley' whose location depends
  on the mass scale at which reionisation affects the cooling of gas
  in haloes ($\vcut$ in our parameterisation).

\item The abundance of satellites fainter than the position of the dip
  is determined by {\it when} reionisation occurred ($\zcut$ in our
  parameterisation); earlier reionisation inhibits the build-up of a
  significant population of faint satellites and vice versa.

\item The abundance of satellites brighter than the position of the
  dip is unaffected by the redshift of reionisation, as these galaxies
  typically assemble the bulk of their stellar mass long after
  reionisation.
\end{enumerate}
In principle, the signatures of reionisation described in (1)-(4) are
measurable.  Remarkably, our general prediction that the satellite
luminosity function is made up of two distinct components seems to be
validated by the observed satellite luminosity function: by combining
a recent estimate for the Milky Way \citep{Newton2017} with an
estimate for M31 based on the PAndAS survey \citep{McConnachie2009},
we find that the presence of a bimodal distribution is preferred over
a simple power law (Fig.~\ref{fig:AICtest}). Although with somewhat
larger uncertainties, we also find that the existence of two
populations can be inferred from the Milky Way data alone
(Fig.~\ref{fig:AICtest_noM31}).

The observed number of Milky Way satellites brighter than ${\rm M}_V
\approx -8$ is well reproduced in both the L16 and H16 \Galform{}
models (Fig.~\ref{fig:cumulativeAll}), but both vastly underpredict
the abundance of galaxies fainter than this magnitude
(Fig.~\ref{fig:differentialAll}). This is because in these models,
$\zcut$ is large, 10 in L16 and 7.9 in H16. With such large values of
$\zcut$, gas cooling shuts off too early, preventing the formation of
faint galaxies after this time.  The large values of $\zcut$ adopted
in these \Galform{} models were chosen by reference to the value of
the redshift of reionisation inferred from early {\it Planck} data,
$z_{re} \sim 11$ \citep{Planck2014}. In the latest {\it Planck} data
analysis, this value has come down to $z_{re}=8.8^{+1.7}_{-1.4}$.

Perhaps not coincidentally, a variant of the L16 model (L16-$z$6)
where $\zcut=6$ provides an excellent match to both the cumulative and
differential versions of the observed luminosity function down to the
faintest magnitudes; in particular it produces many more satellites
fainter than ${\rm M}_V=-7$ than L16, with the faintest satellites
assembling the bulk of their stellar mass prior to reionisation
(Fig.~\ref{fig:reionFrac}). The choice of $\zcut=6$ is also appealing
as it is the self-consistent value for the redshift of reionisation in
the L16 model (see \S\ref{sect:specifics}); in addition, L16-$z$6
retains all the successes on large scales of the L16 model. While
$\zcut=6$ may appear `too late' compared to the latest {\it Planck}
value of $z_{re}$, it should be noted that in the {\it Planck}
analysis $z_{re}$ is defined as the time when the Universe is 50\%
ionised. By contrast $\zcut$ is more readily interpreted as the time
when reionisation is complete. Given the large quoted uncertainty in
{\it Planck}'s $z_{re}$, these two values are compatible. Furthermore,
$\zcut=6$ is also consistent with the inference from the absorption
spectra of QSOs that reionisation should have been completed by $z\sim
6$. 

The epoch by which a region of the Universe is completely reionised
depends on its environment \citep[see,
  e.g.,][]{Alvarez2009,Font2011,Dawoodbhoy2018}. Our assumed value of
$\zcut=6$, which results in a good match to the luminosity functions
of the Milky Way and M31, could differ for galaxies located in regions
of higher or lower overdensities due to the presence of a larger or
smaller population of local ionising sources. While this would affect
the number of galaxies predicted to have formed prior to reionisation,
even relatively large changes to $\zcut$ have very little impact on
the {\it scale} of the transition between the pre- and
post-reionisation population of satellites
(c.f. Fig.~\ref{fig:zcutVars}).

Finally, we have predicted the number of satellites expected to be
present around galaxies similar in mass to the LMC
(Fig.~\ref{fig:cumulativeLMC}). The L16-$z$6 model, which provides the
best match to the combined Milky Way and M31 satellite data, predicts
$26 \pm 10$ satellites (68 per cent confidence) brighter than ${\rm
  M}_V=0$. As shown in Fig.~\ref{fig:pdfLMC}, the majority of the
contribution to this population is from galaxies with ${\rm M}_V \sim
-3$, or $M_\star \approx 10^3\,M_\odot$.

With the continuing investment in observational efforts to compile a
census of satellites around galaxies other than our own \citep[see,
  e.g. the recent results from the SAGA survey;][]{Geha2017}, the
statistical significance of the features detected in the satellite
luminosity function of the Milky Way and M31 may be confirmed. The
prospect of also detecting satellites around less massive galaxies,
such as the LMC, offers the possibility of a further test of current
ideas about some of the most fundamental physical processes involved
in galaxy formation.

\section*{Acknowledgements}
We thank the anonymous referee for useful suggestions that improved
our paper and Marius Cautun and Andrew Cooper for valuable comments on
an early draft. We are also grateful to Alan McConnachie for useful
comments on the completeness of the PAndAS catalogue, and for making
it publicly available in the first place. SB is supported by Harvard
University through the ITC Fellowship. AJD is supported by a Royal
Society University Research Fellowship, and she and CSF by the STFC
Consolidated Grant for Astronomy at Durham (ST/L00075X/1). This work
used the DiRAC Data Centric system at Durham University, operated by
the Institute for Computational Cosmology on behalf of the STFC DiRAC
HPC Facility (\url{www.dirac.ac.uk}). This equipment was funded by BIS
National E-infrastructure capital grant ST/K00042X/1, STFC capital
grants ST/H008519/1 and ST/K00087X/1, STFC DiRAC Operations grant
ST/K003267/1 and Durham University. DiRAC is part of the National
E-Infrastructure. This research was carried out with the support of
the HPC Infrastructure for Grand Challenges of Science and Engineering
Project, co-financed by the European Regional Development Fund under
the Innovative Economy Operational Programme. The data analysed in
this paper can be made available upon request to the author.

\bibliographystyle{apj}
\bibliography{lf}{}

\end{document}